\documentstyle[12pt]{article} \textheight=8.5in
\begin{document} \baselineskip=16pt

\def\FramedPicture(#1 wide #2){ \dimen0= #1
\vcenter{} } \def\eqn#1{\[#1\]}
\newcommand\beginpaper{\end{abstract}\end{titlepage}}
\def\npb#1#2#3{Nucl. Phys. {\bf#1} (#2) #3}
\def\plb#1#2#3{Phys. Lett. {\bf#1} (#2) #3}
\def\prd#1#2#3{Phys. Rev. {\bf#1} (#2) #3} \def\prl#1#2#3{Phys.
Rev. Lett. {\bf#1} (#2) #3} \def\prt#1#2#3{Phys. Rep. {\bf#1}
(#2) #3} \def\zph#1#2#3{Z. Phys. {\bf#1} (#2) #3}
\def\mpl#1#2#3{Mod. Phys. Lett. {\bf#1} (#2) #3}
\def\PubNumber#1{\begin{titlepage}\begin{flushright} UM--TH--#1
\end{flushright}} \def\title#1{\vskip 3cm \begin{center}
{\bf\large #1} \end{center}} \def\andtitle#1{\vskip 4pt
\begin{center}{\bf\large #1} \end{center}}
\def\author#1#2{\vskip 1cm\begin{center}{\bf #1\\} \vskip 2pt
{\it #2}\end{center}} \def\andauthor#1#2{\vskip 0.25cm {and}
\vskip 0.25cm \begin{center}{\bf #1\\}\vskip 2pt {\it #2}
\end{center}} \newcommand\beginabstract{\vskip
2cm\begin{abstract}}
\def\cent#1#2{\setbox0=\hbox{$#1$}\setbox1=\hbox{#2}
{#1\kern-.5\wd0\kern-.5\wd1#2\kern.5\wd0}}
\def\slash#1{\setbox0=\hbox{$#1$}\setbox1=\hbox{$/$}#1
\kern-.4\wd0\kern-.5\wd1{\hbox{\raise.4pt\hbox{/}}}
\kern-.5\wd1\kern.4\wd0}
\def\cut{\hfill\break\noindent}\def\det{{\rm det}}
\def\adhoc{{\it \ ad hoc\ }} \def\backhalf{\kern-2pt}
\def\backspace{\kern-6pt} \def\brk{\hfill\break} \def\etal{{\it
et al.}} \def\ie{{\it i.e.}} \def\ibid{{\it Ibid.}}
\def\ev{{\rm e\backhalf V}} \def\gev{{\rm Ge\backhalf V}}
\def\mev{{\rm Me\backhalf V}} \def\tev{{\rm Te\backhalf V}}

\def\calA{{\cal A}}\def\calB{{\cal B}} \def\calC{{\cal
C}}\def\calD{{\cal D}} \def\calE{{\cal E}}\def\calF{{\cal F}}
\def\calG{{\cal G}}\def\calH{{\cal H}} \def\calI{{\cal
I}}\def\calJ{{\cal J}} \def\calK{{\cal K}}\def\calL{{\cal L}}
\def\calM{{\cal M}}\def\calN{{\cal N}} \def\calO{{\cal
O}}\def\calP{{\cal P}} \def\calQ{{\cal Q}}\def\calR{{\cal R}}
\def\calS{{\cal S}}\def\calT{{\cal T}} \def\calU{{\cal
U}}\def\calV{{\cal V}} \def\calW{{\cal W}}\def\calX{{\cal X}}
\def\calY{{\cal Y}}\def\calZ{{\cal Z}}

\def\newpage{\vfil\break} \def\vec#1{{\overrightarrow {#1}} }
\def\cev#1{{\overleftarrow {#1}} } \def\proc{$q\bar
q\rightarrow Z+{\rm jet}$} \def\C{{\rm C}} \def\out{{\rm out}}
\def\in{{\rm in}} \def\tree{{\rm tree}} \def\con{{\rm con.}}
\def\pslrarw{{\slash{\overrightarrow\partial}}}
\def\psllarw{{\slash{\overleftarrow\partial}}}
\def\ppsi{{\psi^\prime}} \def\pq{{q^\prime}} \def\b{\bar}
\def\qb{\bar q} \def\d#1{{\dot#1}} \def\gm{\gamma_5}
\def\wp{W^+} \def\wm{W^-} \def\z{Z^0} \def\bb{\bar b}
\def\tb{\bar t} \def\top{$t$--quark} \def\bottom{$b$--quark}
\def\h{h} \def\gev{{\rm GeV}} \def\cth{\cos\theta}
\def\sth{\sin\theta} \def\cph{\cos\phi} \def\sph{\sin\phi}
\def\s{\hat s} \def\sp{\hat s^\prime} \def\t{\hat t}
\def\tp{\hat t^\prime} \def\u{\hat u} \def\up{\hat u^\prime}
\PubNumber{92--27} \title{Could Large CP Violation Be Detected
at Colliders?} \author{C. J.--C. Im, G. L. Kane, and P. J.
Malde} {Randall Laboratory of Physics\\University of Michigan,
Ann Arbor, MI 48109\\} \beginabstract We argue that
CP--violation effects below a few$\times 10^{-3}$ are probably
undetectable at hadron and electron colliders. Thus only
operators whose contributions interfere with tree--level
Standard Model amplitudes are detectable. We list these
operators for Standard Model external particles and some two
and three body final state reactions that could show detectable
effects. These could test electroweak baryogenesis scenarios.
\beginpaper \section{Introduction} Our understanding of the
baryon asymmetry of the universe is at an exciting stage of
development. Ideas that show promise for explaining the baryon
asymmetry at the electroweak scaleare being studied\cite{BG}.
So far, all such approaches require large CP--violation, \ie,
CP--violating terms in the Lagrangian with coefficients of the
same order as the gauge couplings.

If such terms exist, their presence may be directly detectable
in collisions at the electroweak scale. The purpose of this
paper is to emphasize several processes that can be studied at
present and future colliders to search for large CP--violating
effects, with emphasis on FNAL. While the possibility of
relating such effects to the origin of the baryon asymmetry is
particularly exciting, motivation for studying such processes
is also provided by the simple observation that at the present
time published limits do not exist for the size of most
CP--violating processes at the $100~\gev$ scale. Thus
heretofore undetected large($\sim50\%$) CP--violation could
occur in some processes at high energy hadron colliders.

Existing electroweak baryogenesis scenarios often depend on
CP--violating Higgs interactions such as $i h\bar t \gamma_5
t$. These are probably the most important vertices to study.
Although the motivation for hypothesizing other vertices is
less compelling, given the speculative nature of present
electroweak baryogenesis scenarios we believe that a systematic
study of all processes which could show a large CP--violation
is appropriate.

We understand, of course, that none of the reactions we list
will be easy to study, but we think it will eventually be
possible to carry out such analyses. The implications of a
positive result are large enough that the effort is justified.

We will parameterize the general CP--violation in terms of
CP--violating operators of dimension less than or equal to six.
Some of these vertices occur in various baryogenesis scenarios.
In this paper, we confine our attention to those operators that
involve only the Standard Model (SM) fields; perhaps eventually
operators involving, for example, superpartners can be studied.

CP--violation parameters at low energies such as $\epsilon$,
$\epsilon^\prime$, and $d_n$\cite{edmN} generally place only
weak constraints on higher dimensional CP--violating operators
because some of these operators contain derivative couplings
that provide a factor of $\hat s$ that leads to suppression at
low energies. At collider energies, however, these operators
can be as large as the SM vertices. At the present time we have
only made qualitative analyses of such constraints and checked
that none of the processes we examine are excluded from
occurring at significant levels; we will report a more careful
and systematic analysis in the future.

\section{CP Violation at Colliders} A number of analyses of
possible CP--violation effects at colliders have been
published~[3--8]. Some have emphasized the possible role of the
top quark[6--13]. There is, however, a major constraint that we
feel has not been considered sufficiently. For both theoretical
and experimental reasons, we think that it is probably
impossible to detect CP--violation effects of order $10^{-3}$
in collider experiments.

The first reason is that the detectors will not be
CP--invariant. Systematic studies can be done to determine at
what level asymmetries in electric or magnetic field lines,
nonuniformity in acceptance efficiency, or spatial asymmetries
in the detector could induce an apparent CP--asymmetry.
Intuitively one might guess they could be of order $10^{-3}$.
To argue they were smaller than that level would require
careful studies of SM processes that are not sensitive to
CP--violation effects. In this experimental ``proof,'' it will
be necessary to get the errors on charge and parity dependent
measurements of particular processes below $0.1\%$. This could
be very difficult, since even the most abundant process that
might allow such a measurement, probably single $W$ production,
will have statistical errors on any measurement even at the SSC
that are of order $10^{-3}$. Whether systematic errors can be
reduced to that level is not known. Furthermore, all analysis
cuts and whatever processes are used to calibrate the detector
must be shown to be CP--invariant at the relevant level.

The second reason is that it will probably be very difficult to
isolate and eliminate spurious CP--violating effects from the
SM processes at the $10^{-3}$ level. Whenever one is studying
CP--violation by actually studying ``naive T''--violation and
assuming CPT--invariance, one has to be sure that spurious
``T"--violating effects such as final state
interactions\cite{Jiang,EHS} are not present. For example,
gluon exchange induces an apparent parity--violating transverse
polarization of order $1\%\sim 2\%$ in $t \bar t$
production\cite{KLY}. This effect can be approximately
calculated\cite{KLY} and a correction made both theoretically
and experimentally, but it will be difficult to eliminate a
residual effect of order $0.2\%$. The process $u\bar d \to
t\bar b$ provides another example. For this reaction, one can
search for CP--violation by studying $u\bar d \to t\bar b$ and
looking for ``T"--violating observables formed from momenta and
the top spin. Then final state QCD interactions and top width
effects both induce such observables in the range
$0.1$---$1\%$. Yet another example is the $Wjj$--channel,
perhaps plus softer jets, that will be a background for $t\bar
t$. Parton level processes in which quarks scatter by
exchanging a gluon and one of the quarks radiates a $W$ will
interfere with processes in which the quarks scatter by
exchanging a $Z$ and one of the quarks or the $Z$ radiates a
$W$, generating an irreducible parity--violating component in
the background; this can easily be of order 1\%. Perhaps it can
be reduced by an appropriate choice of bins and cuts. The
difficulty is that one must find all such effects and eliminate
them before one could believe that there is a new source of
CP--violation.

Furthermore, even when comparing CP conjugate reactions
corrections must be made for structure function differences and
backgrounds. These effects are partly measurable and calculable
in SM, so they can be partly corrected for, but it would take a
great deal of effort and confidence to believe in a new effect
that was much below about a few tenths of a percent. In
addition, any effects that depend on top spin may be affected
by some hadronization of the top quark that polarizes or
depolarizes the top quark spin.

Even if an observable that does not suffer from these effects
could be constructed, it cannot signal CP--violation on
event--by--event basis. Thus, in order to probe CP--violating
effects at the $10^{-3}$ level, it would require at least
$10^{6\sim 7}$ events of some particular type. This is already
at the limit of capabilities of SSC/LHC.

Because of these arguments we conclude that it is probably
impossible to establish new CP--violating effects of order
$10^{-3}$ at colliders. That is not much of a constraint on
FNAL searches, where most channels will have statistical limits
of the same order or larger, but it may limit searches at
SSC/LHC where statistical effects might approach the
$10^{-3}$---$10^{-4}$ level.

There is a qualitative difference in the physics one can study
with $10^{-2}$ effects and with $10^{-3}$ effects. Since
CP--violating effects always arise from interferences, and
since all loops in the SM are already suppressed by factors of
order $10^{-3}$, if $10^{-3}$ is indeed a lower limit on what
could be discovered, we conclude that {\it only new
CP--violating effects that interfere with SM tree amplitudes
could be detected at colliders}.

Given this conclusion, we can enumerate all processes in which
new CP--violating effects could be observed. It can be shown
that we only need to consider processes that involve at least
one top quark or boson--boson couplings (such as gauge
self--couplings or Higgs--gauge boson couplings). As a
corollary, we find that even if there is a CP--violating effect
in the process $p\bar p\rightarrow Z g$, it will be
unobservable at collider experiments, contrary to a recent
speculation\cite{Nachtmann}. This point will be elaborated
elsewhere.

Some of the processes that exhibit tree--level CP--violation
are shown in Table 1 and Table 2. Fortunately all interesting
vertices in the SM are present, though very large luminosity
would be required to study them all down to the $10^{-3}$
level. In the second column of Table 1, we only show a typical
hypothetical CP-violating diagram. These interfere with the
CP--even contribution from the SM to yield tree--level
CP--violation effects. Not all of the processes generated this
way yield an observable in practice. In $gg\to g\to t\bar t$
with a CP--violating $ggg$--vertex, for example, CP--violating
effect vanishes upon averaging over initial gluon spins. In the
third column we show CP-violating operators that correspond to
the CP--odd diagrams. We have written them in a transparent
form, but in actual calculations we use operators that are
fully gauge--invariant\cite{BR,Rujula,Einhorn}. Thus,
$\partial_\nu$ really is the covariant derivative $D_\nu =
\partial_\nu - i g_2 W_\nu^a \tau^a/2 - i g_1 B_\nu/2 - i g_3
G_\nu^a \lambda^a/2$ which connects $W\bar t b$ vertex to
$G\bar t t$ vertex, etc.

Since some of the CP--violating operators are dimension 6, they
have an effective coefficient proportional to $\Lambda^{-2}$,
where $\Lambda$ is some mass scale characteristic of the new
physics. Then the contribution to observables could have a
factor of $\hat s/\Lambda^2$, and the effect will grow with
energy. At $e^+e^-$ colliders this will be a useful effect, but
at hadron colliders the structure functions cut off such an
enhancement.

\section{Observables} In general, there are two ways to observe
CP--violation in high energy processes[3--8]. One can compare
CP--conjugate reactions, such as $b W^+\rightarrow th$ and
$\bar b W^-\rightarrow \bar t h$ at the appropriate angles.
Then it is necessary that electric charges and certain
kinematic quantities be measured. In most cases that will
eventually be possible (see sec 3.4). Alternatively, assuming
CPT--invariance, one can look for ``T''--violating observables
in a single process as long as we look for effects larger than
the expected final state interaction (FSI). The sensitivity of
this method can be sharpened somewhat by calculating the
expected FSI.

In two body reactions one needs a spin as well as momenta to
form CP--violating observables. For top--quark production
processes, the simplest ``T''--violating observable is
$\calO_1= \hat \sigma_t \cdot\hat n$, where $\hat \sigma_t$ and
$\hat n$ are the top--spin and the unit vector normal to the
top production plane. The top--spin can be analyzed
unambiguously by letting the top decay into a $b$--quark and a
$W$ and measuring their momenta, or even from the charged
lepton from the $W$ decay\cite{Kane}. For a top production
process $ab\rightarrow tX$, the corresponding CP--violating
observable is $\hat\sigma_t\cdot
(\vec{p_t}\times\vec{p_a})+\hat \sigma_{\bar
t}\cdot(\vec{p_{\bar t}} \times\vec{p_{\bar a}})$. However, if
the incoming $a$ can be in either of the collider beams, then
averaging over the two possible beam directions makes $\calO_1$
identically zero. In this case, one has to look for a more
complicated ``T''--violating observable that does not vanish
upon averaging over the two possible directions of the incoming
$a$. Finally, one must verify explicitly that the observable
thus constructed yields a non--vanishing expectation value. We
have followed this procedure in this paper.

To convert observables containing the top spin into observables
containing the momenta of the decay products of the top,
replace $\hat \sigma^\mu$ in the observable defined in terms of
the top spin by $q_w^\mu-q_b^\mu(q_w\cdot q_b)/M_w^2$. The
momenta $q_b$ and $q_w$ are the momenta of the decay products
of the top, $p_t=q_w+q_b$. Thus, $\hat \sigma\cdot\hat n =
\hat\sigma\cdot(\vec{p_t} \times \vec{p})$ is equivalent to
$\vec{q_w}\cdot(\vec{q_b}\times \vec{p})$, where $p$ is the
momentum of one of the incoming particles.

\subsection{$bW^+\to th$} Once top and a Higgs boson are
discovered, one can imagine studying this important process.
The Higgs boson will decay to $b\bar b$ with $M_{b\bar b
}=M_h$. There will be electroweak--QCD background with the same
characteristics as signal events, but the background will not
produce a CP--violating effect.

The simplest observable $\hat\sigma_t\cdot(\vec{p_b}
\times\vec{p_t})$ works if we can identify event--by--event the
direction of the incoming $b$--quark. This may be achievable in
practice by exploiting the fact that the energy distribution of
the incoming $W$ in the proton beam is significantly lower than
that of the incoming $b$--quark. Hence the direction of the
center--of--mass momentum of the $th$ system, which is along
one of the beam directions, is highly correlated with the
direction of the initial $b$--quark momentum. Thus the
appropriate observable in this case is
$\vec{p_h}\cdot(\vec{p_t}\times \hat\sigma_t)$.

If $\vec{p_b}$ is not identified event--by--event, the simplest
observable is $\vec{p_t}\cdot\vec{z}~
\vec{z}\cdot(\vec{p_t}\times\hat\sigma_t)$. Equivalently, in
terms of the decay product momenta of the top--quark, the
observable is $\vec{p_t}\cdot\vec{z}~\vec{z}\cdot(\vec{q_b}
\times \vec{q_w})$ (see Table 1). Since the charged lepton from
$t$ semileptonic decay goes preferentially in the direction of
the top spin\cite{Kane}, one can replace $\hat \sigma_t$ by
$\vec{p_{l^+}}$ in any observable. For $\bar t$,
$\hat\sigma_{\bar t}$ should be replaced by $-\vec{p_{l^-}}$.
With a parity--even phase space used in analysis, the SM
predicts vanishing expectation values for these observables.

\subsection{$q\bar q\rightarrow Zh$} The situation here is
similar to that of $bW^+\to th$. The $h$ is only used to
provide a direction, and is detected by selecting $b\bar b$
with $M_{b\bar b}= M_h$. Background from production of
$Z+g(\rightarrow b\bar b)$ will necessarily be present but will
not produce a CP--violation effect. The analysis is similar to
that for $bW^+\to th$, with the $Z$ polarization vector
replacing the top spin direction, and in practice the $Z$
polarization is analyzed by its decay into $l^+l^-$ or into
$q\bar q$\cite{song}.

For $q\bar q\rightarrow Zh$ at FNAL, the initial $q$ and $\bar
q$ carry approximately equal fractions of the beam momentum so
that it is necessary to use observables independent of the
directions of the incoming momenta as written in Table 1. At
$pp$ colliders the $q$ will typically carry a larger fraction
of momenta then the $\bar q$, thus there is a correlation
between $\vec{p_q}$ and the direction of the motion of the
center--of--mass of $Zh$ system. In this case, the simple
observable $(\vec{p_Z}+\vec{p_h})\cdot(\vec{\epsilon_Z}\times
\vec{p_Z})$ can be used, with the momentum of one of the $Z$
decay products replacing $\vec{\epsilon_Z}$ in practice.

\subsection{$gW^+\rightarrow t\bar b$, $u\bar d\rightarrow
t\bar b$, and $ud\rightarrow tb$} The situation here is
analogous to $th$ production. If an effect is ever found, it
will be possible to untangle which process is involved by using
top production rate and decay angular distribution information.
The $\bar b$ (or $b$) provides a direction and possibly a way
to discriminate between $t\bar b$, $\bar t b$, $tb$ final
states. The process $ud\rightarrow tb$ is doubly CKM
suppressed.

\subsection{$q\bar q\rightarrow g q \bar q$ and $gg \rightarrow
ggg$} Construction of an observable for $g(p)g(\bar
p)\rightarrow g(q_1) g(q_2)g(q_3)$ is complicated by the fact
that the simplest observables, such as the triple vector
product, are antisymmetric in its momenta and the cross section
is symmetric. For this process, any observable that depends on
ordering of jets (according to their energies, etc.) will have
exactly vanishing expectation value. This contrasts with the
cases considered by Donoghue and Valencia\cite{DV}. By trial
and error, one finds that the simplest observable that is
symmetric in $p$ and $\bar p$ and in $q_1$, $q_2$, and $q_3$ is
$\calO_g= \hat z\cdot (\vec{q_1}\times \vec{q_2}) \hat
z\cdot(\vec{q_1}-\vec{q_2})\hat z\cdot (\vec{q_1}
-\vec{q_3})\hat z\cdot(\vec{q_2}-\vec{q_3})$, where $\hat z$ is
a unit vector along one of the beam directions.

The ALEPH group\cite{ALEPH} has published results implying that
the electric charge of energetic jets can be measured by
performing an appropriately weighted sum over particles in the
jet, using techniques based on earlier studies of the
JADE\cite{JADE} and MAC\cite{MAC} groups. There appears to be
no reason\cite{GK} why these techniques could not be used at
hadron colliders. Assuming they can be used, to study $q\bar
q\rightarrow gq \bar q$ one should select events with three
jets with one jet having positive electric charge, one
negative, and one zero, summing over all quark types. This
should ensure a sample mainly from $u\bar u\rightarrow g q\bar
q$ and $d\bar d\rightarrow g q \bar q$, separating them from $u
\bar d$ or $uu$ initiated events. At SSC/LHC, the events to
consider are $uu\rightarrow gqq$ and $dd\rightarrow gqq$, so
that one would select events with two like--sign and one
neutral jet. In this case the simplest observable is
$(\vec{q_1}-\vec{q_2})\cdot \hat z
(\vec{q_1}\times\vec{q_2})\cdot\hat z$, where $q_1$ and $q_2$
are the momenta of the jets from the $q\bar q$ pair and $\hat
z$ is along one of the beam directions.

Gluons radiated off quarks will not induce any apparent large
CP--violating effect, but will dilute any real effect, so they
should be suppressed by cutting out events where the neutral
jet is near the beam direction or either of the final quarks.
Final state interactions will cause effects of order
$\alpha_s/\pi$, so only a signal larger than this could be
trusted (see sec. 2). On the other hand, too large an effect
here would induce a neutron electric dipole moment. We estimate
that there is room between these constraints to look for an
effect of order $0.1$. \section{Summary} As discussed in the
introduction, there is good motivation to look for new, large
CP--violation effects at the electroweak scale. Because it is
crucial to demonstrate experimentally that spurious
CP--violation from detector and electroweak--QCD effects
(examples are given in section 2) are absent at the level of
any claimed effect, We have argued that CP--violating effects
of order $10^{-3}$ are probably unobservable at colliders.

Given this conclusion, it is only possible to observe a new,
CP--violating contribution that interferes with a SM tree level
process; any SM loop correction is already too small to be
observable at colliders. Then only a small number of such
processes could be detected. We have listed most processes with
external SM particles that could show such effects, and
described how to analyze data to search for them.

In the future we will report similar analyses for external
supersymmetric partners and perhaps other non--SM particles. We
hope that eventually either such large CP--violating effects
can be detected, or that limits can be obtained that are
relevant to understanding baryogenesis at the electroweak
scale.

\section{ACKNOWLEDGEMENT} C. I. thanks the theory groups at the
Fermi Laboratory, the Brookhaven Laboratory, and Seoul National
University, Korea, for kind hospitality. C. I. also thanks C.
P. Yuan and Tao Han for their help. G. K. appreciates
conversations with A. Dolgov, M. Voloshin, and particularly
with D. Amidei. Both C. I. and G. K. acknowledge valuable
conversations with J.--M. Fr\`ere. This work was supported in
part by the U.S. Department of Energy. \newpage \section{Table
1} In Table 1 we list several two body processes that can be
tested at hadron colliders to detect large CP--violation. For
each process we show one of several diagrams that contribute,
where a solid--circle stands for the CP--violating vertex;
these interfere with the tree--level SM amplitudes of the same
form. The next column gives one of the CP--violating operators
that contribute to the process, and the final column lists ways
to observe the effect. In all cases, $\hat \sigma_t$ is the
top--quark spin, $\hat z$ is one of the beam directions,
$\vec{p_t} (\vec{p_Z})$ is the momentum of the top--quark (Z),
and $\vec{q_+}$ is the momentum of the positively charged decay
product of the $Z$ in $q\bar q\rightarrow Z^0 h$. There are two
possible choices for $\hat z$, but the observables are
independent of this choice. Finally, $N_+$ ($N_-$) refers to
the number of positively charged decay product of either $Z$ or
top--quark emerging above (below) the $x$--$z$ plane, where the
coordinate system is defined so that $\hat z$ is the beam
momentum making an acute angle with $\vec{p_t}$, and
$\vec{p_t}$ lies in the first quadrant of the $x$--$z$ plane.
$\hat y = \hat z\times \hat x$. The coefficient $\Delta$'s
measure the strengths of the CP--violating operators. Note that
$\Delta_{Wt}$, $\Delta_{Zh}$, and $\Delta_{gt}$ have dimensions
of $M^{-2}$, while $\Delta_{ht}$ is dimensionless.
\eqn{\matrix{{\rm Reaction}&{\rm Example}&{\rm \slash{CP}
Operator}&{\rm Observables}\cr u\bar d\rightarrow t\bb&
\FramedPicture(1in wide udtotb.ps)&i\Delta_{Wt}W^+_{\mu\nu}\bar
t_L\gamma^\mu \partial^\nu b_L&{\vec{p_t}\cdot\hat
z\hat\sigma_t\cdot(\vec{p_t}\times \hat z)~~,~~ N_+-N_-}\cr
q\qb\rightarrow\z\h&\FramedPicture(1in wide qqbartozh.ps)
&\Delta_{Zh}\h \epsilon^{\mu\nu\sigma\eta}\tilde
Z_{\mu\nu}Z_{\sigma\eta} & {\vec{p_Z} \cdot\hat
z\vec{q_+}\cdot(\vec{p_Z}\times\hat z)~~,~~ N_+-N_-}\cr
g\wp\rightarrow t\bar b&\FramedPicture(1in wide gwtotb.ps)
&i\Delta_{gt} G^a_{\mu\nu}\bar t\gamma^\mu \lambda^a
\partial^\nu t& { \vec{p_t}\cdot \hat
z\hat\sigma_t\cdot(\vec{p_t}\times\hat z)~~,~~ N_+-N_-}\cr
b\wp\rightarrow t\h&\FramedPicture(1in wide bwtoth.ps)&i
\Delta_{ht}\h\bar t\gm t&{\vec{p_t}\cdot\hat
z\hat\sigma_t\cdot(\vec{p_t} \times\hat z)~~,~~ N_+-N_-}\cr }}
\newpage \section{Table 2} In Table 2 we list some of the three
body processes that can be tested at FNAL to detect large
CP--violation. The entries are defined as in Table 1. In all
cases, $p_i$'s ($q_i$'s) are the incoming (outgoing) momenta.
If the charges of parent partons of jets can be identified (see
text for details), we use the sign of their charges as
subscripts of the corresponding momenta. Hence, $\vec{q_0}$ is
the momentum of a gluon jet, $\vec{q_+}$ the momentum of a $u$,
$\bar d$ or $\bar s$, etc. For $gg\rightarrow ggg$, the
observable $\calO_g$, defined in sec. 3.4, is totally symmetric
in the $q_j$'s and in the $p_i$'s. This observable can also be
used for $qq \rightarrow qqg$. The simplest observable for
$qq\rightarrow qqg$ is symmetric in the two momenta of the jets
coming from the charged quark pairs. Hence, in practice, one
only needs to isolate the neutral jet. The vector $\hat z$ is
along one of the beam directions. \eqn{\matrix{ {\rm
Reaction}&{\rm Example}&{\rm \slash{CP} Operator}& {\rm
Observables} \cr g g\rightarrow g g g&\FramedPicture(0.8in wide
ggtoggg.ps)&\Delta_g
f_{abc}\epsilon^{\mu\nu\sigma\eta}G^{a\rho}_\mu
G^b_{\rho\nu}G^c_{\sigma \eta}&\calO_g\cr q q\rightarrow g q
q&\FramedPicture(1in wide qqtogqq.ps)& \Delta_g f_{abc}
\epsilon^{\mu\nu\sigma\eta}G^{a\rho}_\mu G^b_{\rho\nu}
G^c_{\sigma\eta}&( \vec{q_+}-\vec{q_-})\cdot\hat z
(\vec{q_+}\times\vec{q_-}) \cdot\hat z\cr }}  \end{document}